\documentclass{pasj00}
\draft

\begin{document}
\SetRunningHead{K. Iwai}{Coronal and Chromospheric Magnetic Fields by NoRH}
%\Received{}%{yyyy/mm/dd}
%\Accepted{}%{yyyy/mm/dd}
%\Published{}%{yyyy/mm/dd}

\title{Measurements of Coronal and Chromospheric Magnetic Fields using Polarization Observations by the Nobeyama Radioheliograph}

 \author{
   Kazumasa \textsc{Iwai}\altaffilmark{}
   and
   Kiyoto \textsc{Shibasaki}\altaffilmark{}}
 \altaffiltext{}{Nobeyama Solar Radio Observatory, National Astronomical Observatory of Japan, 
Nobeyama, Minamimaki, Minamisaku, Nagano 384-1305, Japan}
 \email{kazumasa.iwai@nao.ac.jp}

\KeyWords{Sun: chromosphere - Sun: magnetic fields - Sun: radio radiation -@methods: data analysis}

\maketitle

\begin{abstract}
Coronal and chromospheric magnetic fields are derived from polarization and spectral observations of the thermal free-free emission using the Nobeyama Radioheliograph (NoRH). In magnetized plasma, the ordinary and extraordinary modes of free-free emission have different optical depths.  This creates a circularly polarized component in an atmosphere with a temperature gradient. We observed an active region on April 13, 2012 to derive its coronal and chromospheric magnetic fields. The observed degree of circular polarization was between 0.5 \% and 1.7 \%. The radio circular polarization images were compared with ultraviolet images observed by the Atmospheric Imaging Assembly and the photospheric magnetic field observed by the Helioseismic and Magnetic Imager, both on board the Solar Dynamic Observatory. At the edge of the active region, the radio circular polarization was emitted mainly from coronal loops, and the coronal magnetic field was derived to be about 70 G. At the center of the active region, the chromospheric and coronal components cannot be separated. The derived magnetic field is about 20 \% to 50 \% of the corresponding photospheric magnetic field, which is an emission-measure-weighted average of the coronal and chromospheric magnetic fields. 
\end{abstract}

\section{Introduction}
The magnetic field of the solar atmosphere is an important clue to understanding many solar phenomena such as flares and coronal mass ejections. It has been observed in the optical or infrared ranges by using the Zeeman effect at the photosphere. Optical or infrared observations can also be used to measure the coronal and chromospheric magnetic fields using the Zeeman and/or Hanle effect (\cite{tru2005}; \cite{lin2004}), although it is much more difficult because of the weaker magnetic fields in hot, turbulent plasma. The coronal magnetic fields are also derived by extrapolations of the photospheric magnetic fields using the potential field (\cite{sak1982}; \cite{shi2008}) and linear or nonlinear force-free field approximations (e.g., \cite{ino2012}). However, the magnetic field in the photosphere is neither potential nor force-free due to the high gas pressure. Hence, these observation and modeling methods are still being developed and should be verified by comparison with each other. In this study, we derived the chromospheric and coronal magnetic fields by microwave radio observations. 

Several methods are used to derive the magnetic field using radio observations (\cite{gar2004}; \cite{shi2011a} for a review). The magnetic field above sunspots can be derived from the gyro-resonance emission (\cite{dul1985}; \cite{gar1994}). Polarization reversal by quasi-transverse propagation of radio emission is also used for estimating coronal magnetic fields (\cite{coh1960}; \cite{rya1999}).

The longitudinal component of the magnetic field can be derived by thermal bremsstrahlung or so-called thermal free-free emission. \citet{bog1980} observed the radio polarization using one-dimensional scanning spectral observations of RATAN-600. Circularly polarized emission is observed in plage regions at wavelengths between 2.3 and 4.0 cm. The observed circular polarization degrees were up to 1.5 \% and were converted to a coronal magnetic field of 20 to 60 G using the observed radio spectral index.

Observations in a higher frequency range enable us to derive the magnetic field at a deeper layer. \citet{gre2000} observed circular polarization using single-frequency observations with the Nobeyama Radioheliograph (NoRH) at 17 GHz. They used a radiation transfer model simulation to deduce the magnetic field from the observed circular polarization. The ratio of the radio-deduced magnetic field and the corresponding photospheric magnetic field is 0.5 to 0.6. Their radiation transfer model simulation showed that both the coronal and chromospheric components are included in the circular polarization at 17 GHz. 

Microwave radio observation is advantageous because it can be used to derive both the chromospheric and coronal magnetic fields. On the other hand, separation of the chromospheric and coronal components requires spectral observation or modeling. In particular, spectral observation is essential for examining the chromospheric component because the spectral index of the chromosphere is unknown. Hence, the circular polarization observation and spectral observation should be combined (see Equation 1 in Section 2.2). 

The purpose of this study is to derive the chromospheric and coronal magnetic fields by combining two-dimensional radio polarization and radio spectral imaging observations using NoRH at 17 and 34 GHz. Then, the derived radio magnetic fields are evaluated by comparing them with the coronal loop structure and photospheric magnetic field. The instrument and theory to derive magnetic fields are described in Section 2. The data analysis results are presented in Section 3 and discussed in Section 4. 

\section{Observation}
\subsection{Instrument}
NoRH is a radio interferometer dedicated to solar observation; it has 84 antennas, each with a diameter of 80 cm (\cite{nak1994}). NoRH observes the full solar disk every 1 s at 17 GHz (intensity and circular polarization) and 34 GHz (intensity). Radio images are synthesized every 1 s. 

\begin{figure*}
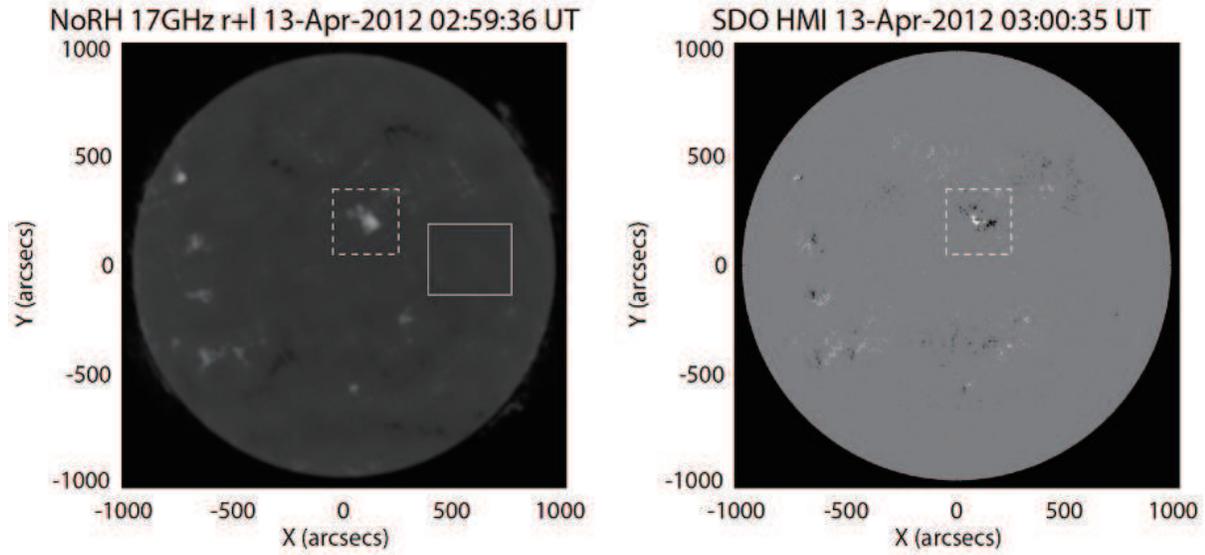

  \begin{center}
    \FigureFile(160mm,80mm){fig1.eps}
  \end{center}
  \caption{(Left) Radio intensity at 17 GHz observed with NoRH at 03:00:15 UT on April 13, 2012. Solid rectangle shows radio-quiet region described in Table 1. (Right) Photospheric magnetic field observed with SDO/HMI at 03:00:35 UT. Dotted rectangles in right and left panels show active region described in Figure 2.}\label{fig1}
\end{figure*}

An active region observed on April 13, 2012 (NOAA 11455), is analyzed in this study. Figure \ref{fig1} shows the radio intensity image observed with NoRH at 17 GHz and the magnetic field observed with the Helioseismic and Magnetic Imager (HMI: \cite{sch2012}) on board the Solar Dynamics Observatory (SDO) on April 13, 2012. There was a small active region near the disk center (N06W19). This situation is suitable for observing the longitudinal component of the magnetic field. 

Three main emission processes which produce microwave solar radio emissions are; free-free emission, gyro-resonance emission, and gyro-synchrotron emission. Gyro-resonance emission is emitted from sunspots at lower harmonics (2 or 3) of their local gyro-frequency.  The third harmonic of the gyro-frequency at 2000 G is about 17 GHz. In Figure \ref{fig1}, the largest magnetic field in the active region observed with HMI is less than 2000 G. Hence, there is no gyro-resonance emission component from the observed active region. Gyro-synchrotron emission is emitted from non-thermal electrons during flares. In this study, flare-quiet periods are selected to avoid contamination by gyro-synchrotron emission. Therefore, the radio emission observed from the active region in this study is purely thermal free-free emission. 

\subsection{Derivation of the magnetic field}
In magnetized plasma, the ordinary and extraordinary modes of free-free emission have different optical depths. That makes a circular polarized component which is inverted to obtain the longitudinal component of the magnetic field $B_l$ as follows (\cite{bog1980}):
\begin{eqnarray}
B_{l}[G] & = & \frac{10700}{n\lambda[cm]}\frac{V}{I} \nonumber \\
n& \equiv & \frac{d(\log I)}{d(\log\lambda)}
\label{eq1}
\end{eqnarray}
where $I$ is the brightness temperature, $V$ is the brightness  temperature of the circularly polarized component, $\lambda$ is the wavelength in cm, and $n$ is the power-law spectral index of the brightness temperature. For an optically thin case, the spectral index is close to 2. This method is used to derive the magnetic fields of coronal loops in limb observations (\cite{shi2011b}).

In the microwave range, the opacity of the free-free emission becomes thick around the upper chromosphere. Equation \ref{eq1} shows that an optically thick region of uniform temperature $(n\approx 0)$ cannot produce the circularly polarized component regardless of the existence of the magnetic field. However, a difference in opacity between the ordinary and extraordinary modes means that these two modes can penetrate into different layers. If a temperature gradient exists between these two layers, the brightness temperatures of the ordinary and extraordinary modes differ. Therefore, the magnetic field in the chromosphere in the presence of a temperature gradient produces a nonzero spectral index and hence also produces the circularly polarized component.

\section{Data Analysis}
\subsection{Data Accumulation Time and Noise Level}
Because the circular polarization degree of free-free emission is very small, hundreds to thousands of synthesized images should be averaged to reduce the noise level of the images. In this study, the averaging period is determined by the standard deviation of the polarized signal in the quiet region.  
The Stokes parameters of $I$ and $V$ in this study are given by,
\begin{eqnarray}
V=(R - L)/2 \nonumber \\
I=(R + L)/2
\label{eqstok}
\end{eqnarray}
where, $R$ and $L$ are the brightness temperatures of the right- and left-handed circular polarized component, respectively. A degree of polarization $(P)$ is given by,
\begin{equation}
P = \frac{V}{I} = \frac{R-L}{R+L}
\label{eupol}
\end{equation}
The standard deviation of $P$ $(=\sigma_{P})$ is given by the law of propagation of errors as follows,
\begin{equation}
\sigma_{P}^{2} = \left(\frac{\partial P}{\partial R}\right) ^2 \sigma_{R}^{2} + \left(\frac{\partial P}{\partial L}\right) ^2 \sigma_{L}^{2}
\label{eqspol}
\end{equation}
where, $\sigma_{R}$ and $\sigma_{L}$ are the standard deviations of R and L, respectively. Equation \ref{eqspol} is solved as follows,
\begin{equation}
\sigma_{P}^{2} = \left(\frac{2L_0}{(R_0+L_0)^2}\right) ^2 \sigma_{R}^{2} + \left(\frac{-2R_0}{(R_0+L_0)^2}\right) ^2 \sigma_{L}^{2}
\label{eqspol2}
\end{equation}
where, $R_0$ and $L_0$ are the averages of R and L, respectively. We assume $R_0 = L_0$ and $\sigma_{R} = \sigma_{L}$ for non-polarized emission. Then, $\sigma_{P}$ is derived as follows,
\begin{equation}
\sigma_{P} = \frac{\sigma_R}{\sqrt[]{2}R_0}
\label{eqsigp}
\end{equation}

The standard deviation of $V$ $(=\sigma_{V})$ is also given by the law of propagation of errors as follows,
\begin{equation}
\sigma_{V}^{2} = \left(\frac{\partial V}{\partial R}\right) ^2 \sigma_{R}^{2} + \left(\frac{\partial V}{\partial L}\right) ^2 \sigma_{L}^{2} = \frac{\sigma_R^2}{2}
\label{eqsigv}
\end{equation}
Hence, $\sigma_{P}$ is derived by Equations \ref{eqstok}, \ref{eqsigp}, \ref{eqsigv} as follows,
\begin{equation}
\sigma_{P} = \frac{\sigma_V}{I_0}
\label{eqsol}
\end{equation}
where $I_0$ is the average intensity, and $I_0 = R_0 = L_0$ for non-polarized emission.

The white rectangle in Figure \ref{fig1} shows the radio-quiet region used in this study, and Table 1 shows the standard deviation of the circularly polarized signal $(\sigma_V)$ in this region. The standard deviations of the polarization component were 21 K after the images were averaged over 2 min and 11 K after averaging over 20 min. The average intensity $(I_0)$ of the solar disk is about 10,000 K. We define five-sigma as the minimum detectable signal level. Hence, the minimum detectable level of a degree of polarization corresponds to 1.0 \% and 0.5 \% for 2 min and 20 min averaging, respectively. Synthesized images are averaged for 20 min in the following data analysis to derive weaker magnetic fields. The solar rotation during 20 min is about 3 arcsec, which is smaller than the beam size of NoRH (10 arcsec at 17 GHz). Hence, the solar rotation effect is neglected in the averaged data.

\subsection{Radio Polarization, Spectra, and Magnetic Field}
Figure \ref{fig2}a shows a radio circular polarization map superimposed on the optical magnetogram observed with HMI on April 13, 2012. The red and blue contours show the positive and negative components of the radio circular polarization, respectively. The location and polarity of the radio circular polarization correspond to those of the optical magnetogram. The degree of polarization is up to 1.7 \% for negative polarity. Figure \ref{fig2}c shows an EUV image at 304 {\AA} observed by the Atmospheric Imaging Assembly (AIA: \cite{lem2012}) on board the SDO. The white contours show the radio intensity at 17 GHz. The radio intensity corresponds to the bright region at 304 {\AA}.

The calibration sequence of NoRH image synthesis uses the sky and the quiet Sun levels. A histogram of the pixel counts included in a synthesized image is derived for each image. The most frequent count level is defined as the background sky level, which is assumed to be at 0 K because the cosmic microwave background radiation is negligible compared with the radiation from the Sun. The second most frequent count level is defined as the quiet Sun level. The brightness temperature of the quiet Sun at microwave to millimeter wavelengths has been observed and modeled in several studies (\cite{lin1973}; \cite{bec1973}; \cite{kus1976}; \cite{zir1991}; \cite{sel2005}). From these studies, the brightness temperatures of the quiet Sun at 17 GHz and 34 GHz are 10,000 K and 9000 K, respectively. The radio spectral index of the quiet region between 17 and 34 GHz is about 0.15 using this model. A linear approximation between the background sky and the quiet Sun enables us to calibrate the observed brightness temperatures. The spectral index of the quiet Sun at microwave range is assumed to be constant (e.g \cite{sel2005}). Hence, we define that the spectral index between 17 and 34 GHz is same as the local spectral index at 17 GHz. The green contours in Figure 2b show the radio spectral index of the active region, which is between about 0.4 and 0.6 around the active region.

The chromospheric temperature structure of active regions might be different from that of the quiet region. Thus, the magnetic field that is derived from the spectral index between 17 and 34 GHz contains an error. Although it is difficult to estimate the extent of this error, it is unlikely that the spectral index of the active region at 17 GHz is far from that of between 17 and 34 GHz. 

\subsection{Magnetic Fields in the Photosphere and Chromosphere}
Figure \ref{fig2}b shows the magnetic field strength derived by substituting the radio circular polarization, intensity, and radio spectral index in Equation \ref{eq1}. The radio magnetic field strength is derived only in regions that have circular polarization degrees of greater than 0.5 \% and spectral index of greater than 0.15. Red and blue contours show the positive and negative components of the radio magnetic field, respectively. The location of the radio magnetic field corresponds to that of the photospheric magnetic field within the beam size of NoRH ($\sim$10 arcsec at 17 GHz). 

Optical magnetograms usually have higher spatial resolutions than the beam size of NoRH. Hence, the area-averaged magnetic fields within 20 arcsec square regions are compared. The radio magnetic fields at the center of the positive and negative polarity regions [footpoint (FP) regions in Figure \ref{fig2}a] are 116 G and -217 G, respectively. The corresponding photospheric magnetic fields are 568 G and -456 G in the positive and negative polarity regions, respectively. The ratio between the photospheric and radio magnetic fields is 0.20 in the positive polarity region and 0.47 in the negative polarity region. 

\section{Interpretation and Discussion}
\subsection{Coronal and Chromospheric Components of the brightness temperature}
The microwaves observed in this study (17 and 34 GHz) are emitted mainly from the chromosphere. However, the coronal component exists especially in coronal loops around active regions. For simplicity, we adopt a two-component atmosphere model consisting of the corona and chromosphere (\cite{zir1991}; \cite{gre2000}). In this model, the observed brightness temperature $(T_{b})$ is given by 
\begin{equation}
T_{b}=T_{e,chr}(\lambda)\exp(-\tau_{c}(\lambda))+T_{e,cor}(\lambda)(1-\exp(-\tau_{c}(\lambda)))
\label{eq2}
\end{equation}
where $\tau_{c}(\lambda)$ is the optical depth of the corona at a given wavelength, and $T_{e,cor}$ and $T_{e,chr}$ are the electron temperatures of the corona and chromosphere, respectively.

The corona and chromosphere are assumed to be optically thin and thick, respectively. In addition, the electron temperature of the corona $(\sim 10^{6})$ is about 100 times larger than that of the chromosphere $(\sim 10^{4})$. Hence, Equation \ref{eq2} is simplified as follows (\cite{gre2000}):
\begin{equation}
T_{b,obs}(\lambda)=T_{b,chr}(\lambda)+T_{b,cor}(\lambda)
\label{eq3}
\end{equation}
where $T_{b,obs}(\lambda)$ is the observed radio intensity, and $T_{b,chr}(\lambda)$ and $T_{b,cor}(\lambda)$ are the chromospheric and coronal components, respectively. In this study, the brightness temperature of the quiet Sun is assumed to be 10,000 K at 17 GHz and 9,000 K at 34 GHz. These can be considered as the base brightness temperature of the chromosphere. Hence, the increment relative to this chromospheric component is estimated to be the coronal component if we neglect the temperature variation in the chromosphere. 

The assumption in Equation \ref{eq3} can be checked by plugging the observational results into Equation \ref{eq2}. In the FP- region in Figure \ref{fig2}a, for example, the observed total radio intensity at 17 GHz $T_{b,obs}(1.76 cm)$ is about 14,100 K. The coronal and chromospheric electron temperatures $T_{b,cor}(1.76 cm)$ and $T_{b,chr}(1.76 cm)$ are assumed to be $10^6$ and $10^4$ K, respectively. These observational results and assumptions yield $\exp(-\tau_{c}(1.76 cm)) \approx 0.99$ or $\tau_{c}(1.76 cm) \approx 0$. Hence, Equation \ref{eq3} is approximately accurate when $\tau_{c}(\lambda) \ll 1$ and $ T_{b,chr}(\lambda) \ll T_{b,cor}(\lambda)$, which are usually true in the solar atmosphere.

\subsection{Coronal and Chromospheric Components of the polarized emission}
The polarized component $V_{obs}(\lambda)$ is also derived as $T_{b,obs}(\lambda)$ is, by using a simple model with a constant coronal magnetic field,
\begin{equation}
V_{obs}(\lambda) = V_{chr}(\lambda) + V_{cor}(\lambda)
\label{eq4}
\end{equation}
However, these two components cannot be separated because the coronal and chromospheric magnetic fields are both unknown. 

Figure \ref{fig3} shows the relationship between the radio circular polarization degree at 17 GHz and the photospheric magnetic field observed by HMI. The radio circular polarizations at the center of the active region (FP+ and FP-) are correlated with the magnetic field in the photosphere. In the EA regions, however, the radio circular polarization is observed even though the corresponding photospheric magnetic fields are almost 0 G. Each area is averaged within 20 arcsec$^2$, which is sufficiently larger than the beam size of NoRH at 17 GHz ($\sim10$ arcsec). Hence, the circularly polarized component at the center of the active region cannot affect the high degree of circular polarization in the EA regions. 

\begin{figure*}
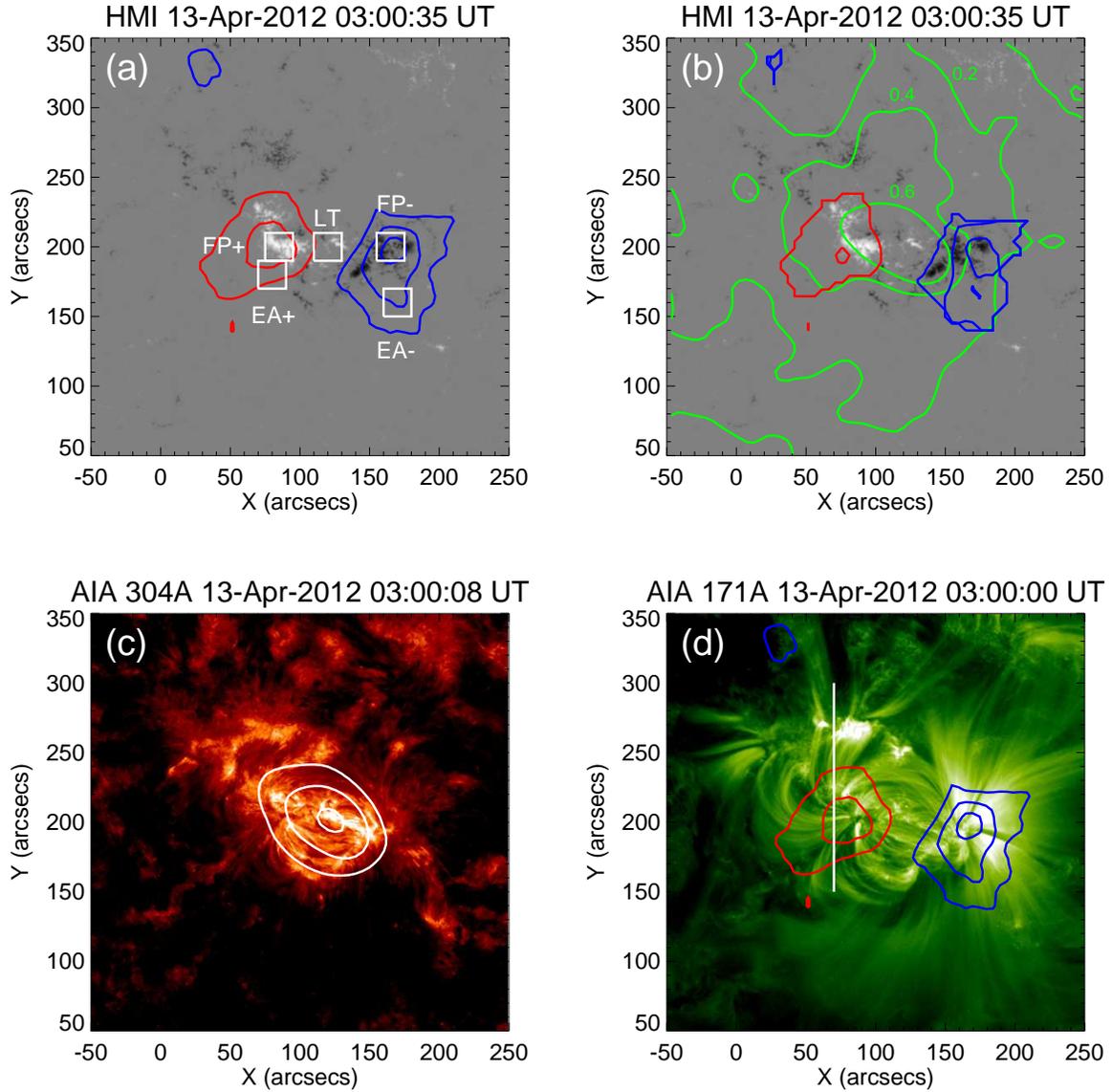

  \begin{center}
    \FigureFile(160mm,160mm){fig2.eps}
  \end{center}
  \caption{(a) Magnetic fields observed by HMI at 03:00 UT on April 13, 2012. Radio circular polarization at 17 GHz is superimposed as contours: positive components in red, 0.5 \%, 1.0 \%; negative components in blue, 0.5 \%, 1.0 \%, 1.5 \%. LT: loop top, FP: footpoint, EA, edge of active region. (b) Magnetic fields observed by HMI at 03:00 UT. Radio magnetic fields at 17 GHz are superimposed as contours: positive components in red, 50, 150 G; negative components in blue, 50, 150, 250 G. Green contours: radio spectral index of the brightness temperature spectrum between 17 and 34 GHz, 0.2, 0.4, 0.6. (c) EUV image at 304 {\AA} observed by AIA. White contours: radio intensity at 17 GHz. (d) EUV image at 171 {\AA} observed by AIA. Red and blue contours indicate radio circular polarization degree at 17 GHz.}\label{fig2}
\end{figure*}

\begin{figure*}
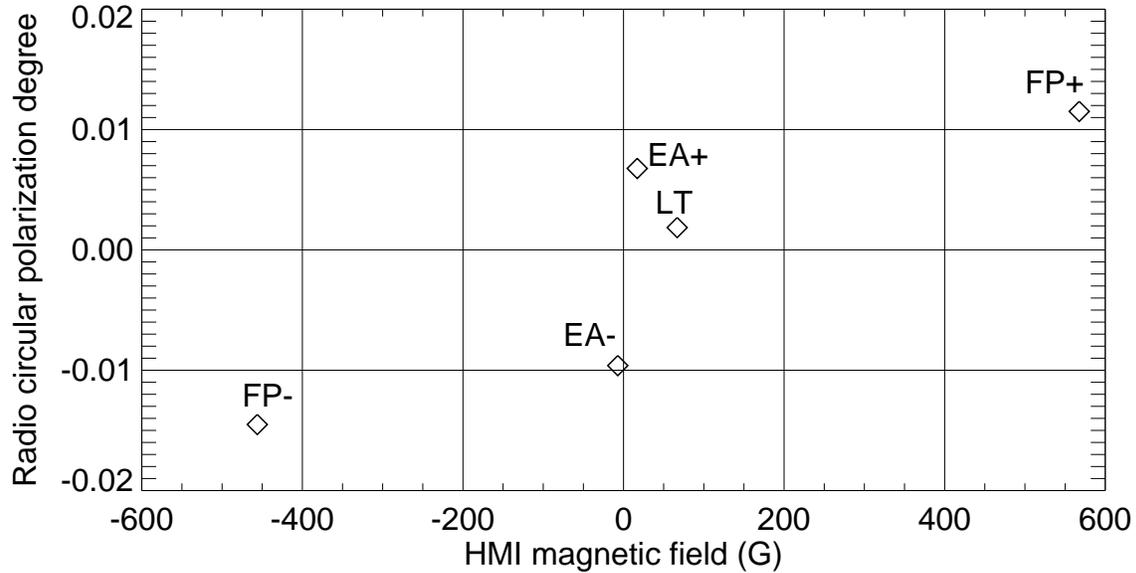

  \begin{center}
    \FigureFile(160mm,80mm){fig3.eps}
  \end{center}
  \caption{Relationship between the radio circular polarization degree at 17 GHz and photospheric magnetic field observed by HMI.}\label{fig3}
\end{figure*}

Figure \ref{fig2}d shows an EUV image at 171 {\AA} observed by AIA. The coronal loop structures exhibit broader structures than the photospheric magnetic field. The red and blue contours show the radio circular polarization degree at 17 GHz. The circularly polarized component is observed between the foot points and loop tops of the coronal EUV loops. In particular, it is clear that the 0.5 \% contour of the positive circular polarization degree lies along the envelope of the tops of the coronal loops. 

\begin{figure}
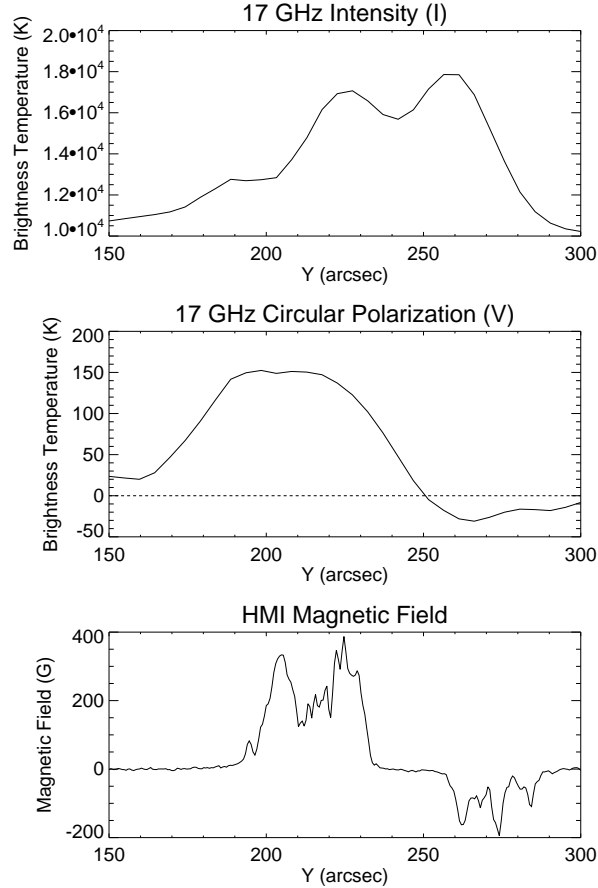

  \begin{center}
    \FigureFile(80mm,160mm){fig4.eps}
  \end{center}
  \caption{(Top) radio intensity at 17 GHz, (middle) circular polarization at 17 GHz, and (bottom) magnetic field observed by HMI, along the white line in Figure 2d.}\label{fig4}
\end{figure}

Figure \ref{fig4} shows the radio intensity, circular polarization, and photospheric magnetic field along the white line in Figure \ref{fig2}d. The circularly polarized component of 17 GHz is observed to the north of 170 arcsec. On the other hand, photospheric magnetic field is observed only north of 195 arcsec. The separation of the two is larger than the beam size of NoRH at 17 GHz ($\sim$ 10 arcsec). The location of the chromospheric magnetic field is thought to be similar to that of the photosphere. Hence, there should be no chromospheric component of circularly polarized radio emission around 170 arcsec. Therefore, the coronal component is the source of the polarized signal, even though the coronal component of intensity is only 1000 to 2000 K as shown in the top panel. 

Now we consider the circular polarization degree in the EA region, which does not contain chromospheric component. In the EA+ region in Figure \ref{fig2}a, the coronal emission is 4040 K, and the circularly polarized component is 96 K. The optically thin coronal component is expected to have a spectral index of 2. These observational values and Equation \ref{eq1} yield a coronal magnetic field of 72 G in the EA+ region. This magnetic field is consistent with the results of \citet{bog1980} that also derived coronal magnetic field using the radio free-free observation at the longer wavelength.

\subsection{Separation of the Chromospheric and the Coronal components}
At the foot points of the active region, the observed circularly polarized signal includes both the chromospheric and coronal components. These two components are emission measure weighted. Because NoRH observes the circular polarization in only one observation band, we cannot separate the circularly polarized component from these two layers. 

Circular polarization observations at multiple frequency bands are effective for distinguishing the chromospheric and coronal circularly polarized components. The use of multiple circular polarization observation bands enables better inversion, especially for coronal three-dimensional magnetic fields. Observations in higher frequency ranges are also effective for reducing contamination by the coronal component and deriving only the chromospheric magnetic field accurately. 

In addition, it is also possible to improve the separation of the chromospheric and coronal magnetic fields by combining radio observations with other observations or models. Optical or infrared observations can be used to measure the chromospheric magnetic fields using the Zeeman or Hanle effect (\cite{tru2005}; \cite{han2005}). The given chromospheric or coronal magnetic fields may enable the separation of the chromospheric and coronal radio circular polarization. 

\section{Summary}
We derived the coronal and chromospheric magnetic fields using circular polarization observations at 17 GHz and spectral observations at 17 and 34 GHz by NoRH. The observational results are summarized as follows:

\begin{itemize}
\item The synthesized images were averaged to reduce their noise level. The minimum detectable level of the circular polarization degree is 1.0 \% and 0.5 \% for 2 min and 20 min accumulations, respectively. The observed circular polarization degree was between 0.5 \% and 1.7 \%.

\item The radio intensity was calibrated using the brightness temperature of the quiet Sun, which is defined as 10,000 K and 9000 K at 17 GHz and 34 GHz, respectively from \citet{sel2005}. The spectral index of the brightness temperature in the quiet region is about 0.15 using this model. The observed spectral index is between 0.4 and 0.6 around the active region. 

\item The magnetic fields are derived from the observed radio circular polarization and the spectral index of the brightness temperature. The ratio of the observed radio magnetic field and the corresponding photospheric magnetic field is about 0.2 to 0.5 at the center of the active region. 
\end{itemize}

The observed radio magnetic fields contain both the coronal and chromospheric components. We assume that the solar atmosphere observed in the microwave range has two components: the optically thin corona and the optically thick chromosphere. The radio circular polarization images were compared with EUV images observed by AIA and the photospheric magnetic field observed by HMI. The results are summarized as follows:

\begin{itemize}
\item At the edge of the active region, radio circular polarization is observed even though the corresponding magnetic field at the photosphere is almost 0 G. At the same time, the location of the observed radio circular polarization corresponds to that of the coronal loop structures. Therefore, we can assume that there is almost no chromospheric circularly polarized component there and derive the pure coronal magnetic field strength. 

\item At the center of the active region, the chromospheric and coronal components cannot be separated, and the derived magnetic field is emission-measure-weighted. It requires additional information for its separation. 
\end{itemize}

For the next step, it is necessary to develop an instrument for two-dimensional polarization observations with high spatial and spectral resolutions in a higher frequency range. These requirements will be achieved by recent and future radio interferometers such as the SSRT (\cite{les2012}), CSRH (\cite{yan2009}), FASR (\cite{bas2004}), and ALMA. These radio observations will be more useful for cooperation with ground- and space-based magnetic field observations such as Solar-C.

%\newpage

\begin{table}
  \caption{Standard deviation of circularly polarized signal $(V)$ and average radio intensity $(I)$ at 17 GHz inside the white rectangle in Figure 1.}\label{table1}
  \begin{center}
    \begin{tabular}{lll}
      \hline
      Averaging time (min) & 2	& 20 \\
      Standard deviation of $V$ (K) & 21 & 11 \\
      Signal threshold (5 $\sigma$) of $V$ (K) & 105 & 55 \\
      Average of $I$ (K) & 10,448 & 10,452 \\
      \hline
    \end{tabular}
  \end{center}
\end{table}

\bigskip

SDO data are courtesy of NASA/SDO and the AIA and HMI science terms. This work was conducted by a joint research program of the Solar-Terrestrial Environment Laboratory, Nagoya University.


\newpage

\begin{thebibliography}{}
\bibitem[Bastian(2004)]{bas2004} Bastian, T. S. 2004, in Solar and Space Weather Radiophysics (Astrophysics and Space Science Library, Vol. 314; Berlin: Springer), 47
\bibitem[Beckman et al.(1973)]{bec1973} Beckman, J. E., Clark, C. D., \& Ross, J. 1973, \solphys, 31, 319
\bibitem[Bogod \& Gelfreikh(1980)]{bog1980} Bogod, V. M., \& Gelfreikh, G. B. 1980, \solphys, 67, 29
\bibitem[Cohen(1960)]{coh1960} Cohen, M.H., 1960, \apj, 131, 664.
\bibitem[Dulk(1985)]{dul1985} Dulk, G. A., 1985, \araa, 23, 169
\bibitem[Gary \& Hurford(1994)]{gar1994}Gary, D. E., \& Hurford, G. J. 1994, \apj, 420, 903
\bibitem[Gary \& Keller(2004)]{gar2004} Gary, D. E., \& Keller, C. U. (ed.) 2004, Solar and Space Weather Radiophysics: Current Status and Future Developments (Astrophysics and Space Science Library, Vol. 314; Dordrecht: Kluwer)
\bibitem[Grebinskij et al(2000)]{gre2000} Grebinskij, A., Bogod, V., Gelfreikh, G., Urpo, S., Pohjolainen, S., \& Shibasaki, K. 2000, \aaps, 144, 169
\bibitem[Hanoka(2005)]{han2005} Hanoka, Y. 2005, \pasj, 57, 235
\bibitem[Inoue et al.(2012)]{ino2012} Inoue, S., Shiota, D., Yamamoto, T. T., Pandey, V. S., Magara, T., \& Choe, G. S. 2012, \apj, 760, 17
\bibitem[Kuseski \& Swanson(1976)]{kus1976} Kuseski, R. A., \& Swanson, P. N. 1976, \solphys, 48, 41
\bibitem[Lemen et al.(2012)]{lem2012} Lemen, J. R., et al. 2012, \solphys, 275, 17
\bibitem[Lesovoi et al.(2012)]{les2012} Lesovoi, S. V., Altyntsev, A. T., Ivanov, E. F., \& Gubin, A. V. 2012, \solphys, 280, 651
\bibitem[Lin et al.(2004)]{lin2004} Lin, H., Kuhn, J. R., \& Coulter, R. 2004, \apj, 613, L177
\bibitem[Linsky(1973)]{lin1973} Linsky, J. L. 1973, \solphys, 28, 409
\bibitem[Nakajima et al.(1994)]{nak1994} Nakajima, H., et al. 1994, IEEEP, 82, 705
\bibitem[Ryabov et al.(1999)]{rya1999} Ryabov, B. I., Pilyeva, N. A., Alissandrakis, C. E., Shibasaki, K., Bogod, V. M., Garaimov, V. I., \& Gelfreikh, G. B. 1999, \solphys, 185, 157
\bibitem[Sakurai(1982)]{sak1982} Sakurai, T. 1982, \solphys, 76, 301
\bibitem[Scherrer et al.(2012)]{sch2012} Scherrer, P. H., et al. 2012, \solphys, 275, 207
\bibitem[Selhorst et al(2005)]{sel2005} Selhorst, C. L., Silva, A. V. R., \& Costa, J. E. R. 2005, \aap, 433, 365
\bibitem[Shibasaki et al(2011a)]{shi2011a} Shibasaki, K., Alissandrakis, C. E., \& Pohjolainen, S. 2011a, \solphys, 273, 309
\bibitem[Shibasaki et al(2011b)]{shi2011b} Shibasaki, K., Narukage, N., \& Yoshimura, K. 2011b, in ASP Conf. Ser. 437, Solar Polarization 6, eds. J. Kuhn et al. (San Francisco, CA: ASP), 433
\bibitem[Shiota et al(2008)]{shi2008} Shiota, D., Kusano, K., Miyoshi, T., Nishikawa, N., \& Shibata, K. 2008, \jgr, 113, A03S05
\bibitem[Trujillo Bueno et al(2005)]{tru2005} Trujillo Bueno, J., Merenda, L., Centeno, R., Collados, M., \& Landi Degl'Innocenti, E. 2005, \apj, 619, L191
\bibitem[Yan et al.(2009)]{yan2009} Yan, Y. H., et al. 2009, Earth Moon Planets, 104, 97
\bibitem[Zirin et al(1991)]{zir1991} Zirin, H., Baumert, B. M., \& Hurford, G. J. 1991, \apj, 370, 779
\end{thebibliography}
\end{document}